\documentclass[pra,aps,superscriptaddress,nofootinbib,twocolumn]{revtex4-1}

\usepackage{mathrsfs}
\usepackage{amsfonts}
\usepackage{amssymb}
\usepackage{amsmath}
\usepackage{graphicx}
\usepackage[usenames,dvipsnames]{color}
\usepackage[colorlinks=true,citecolor=blue,linkcolor=magenta]{hyperref}
\usepackage{ulem}
\usepackage{lmodern}
\usepackage{amsthm}

\newcommand{\figpath}{.}

\newcommand{\Tr}{\mathrm{Tr}}

\newcommand{\abs}[1]{\vert #1 \vert}

\newcommand{\ket}[1]{\vert{ #1 }\rangle}
\newcommand{\bra}[1]{\langle{ #1 }\vert}
\newcommand{\ketbra}[2]{\vert #1 \rangle \langle #2 \vert}

\newcommand{\st}{\,:\,}

\newcommand{\HSket}[1]{\vert{ #1 }\rangle\rangle}
\newcommand{\HSbra}[1]{\langle\langle{ #1 }\vert}
\newcommand{\HSbraket}[2]{\langle\langle #1 \vert #2 \rangle\rangle}
\newcommand{\HSketbra}[2]{\vert #1 \rangle\rangle \langle\langle #2 \vert}

\newtheorem{theorem}{Theorem}
\newtheorem{lemma}{Lemma}

\begin{document}

\title{Quantum operation of fermionic systems and process tomography using Majorana fermion gates}

\author{Gang Zhang}
\affiliation{College of Physics and Materials Science, Tianjin Normal University, Tianjin
300387, China}

\author{Mingxia Huo}
\affiliation{Department of Physics and Beijing Key Laboratory for Magneto-Photoelectrical Composite and Interface Science, School of Mathematics and Physics, University of Science and Technology Beijing, Beijing 100083, China}

\author{Ying Li}
\email{yli@gscaep.ac.cn}
\affiliation{Graduate School of China Academy of Engineering Physics, Beijing 100193, China}

\begin{abstract}
Quantum tomography is an important tool for the characterisation of quantum operations. In this paper, we present a framework of quantum tomography in fermionic systems. Compared with qubit systems, fermions obey the superselection rule, which sets constraints on states, processes and measurements in a fermionic system. As a result, we can only partly reconstruct an operation that acts on a subset of fermion modes, and the full reconstruction always requires at least one ancillary fermion mode in addition to the subset. We also report a protocol for the full reconstruction based on gates in Majorana fermion quantum computer, including a set of circuits for realising the informationally-complete state preparation and measurement.
\end{abstract}

\maketitle

\section{Introduction}

Majorana fermions are candidates for realising the topological quantum computation~\cite{Nayak2008}. They are zero-energy modes or quasi-particle excitations in systems such as topological superconducting nanowires~\cite{Moore1991, Kitaev2001}. The evidence of Majorana fermions has been observed in the experiment~\cite{Mourik2012}, and proposals for realising the quantum computation operations, e.g.~braiding, have been reported~\cite{Alicea2011}. Using Majorana fermions, we can implement fermionic quantum computation without the cost of encoding fermions in qubits~\cite{Bravyi2002, Li2018}. When the noise that changes the topological charge, i.e.~parity of the particle number, is sufficiently suppressed, the fault-tolerant quantum computation using Majorana fermions is more efficient than using conventional qubits~\cite{Bravyi2006}. For instance, the encoding cost of the surface-code-based fault tolerant quantum computation can be greatly reduced by using Majorana fermions~\cite{Li2016}.

In this paper, we develop a framework for the quantum tomography of fermionic systems. Quantum tomography is an important tool for the verification and characterisation of quantum operations~\cite{Chuang1997, Poyatos1997, DAriano2001, Altepeter2003, Mohseni2006, Merkel2013, BlumeKohout2013, Stark2014, Greenbaum2015, BlumeKohout2017, Sugiyama2018}. The tomography can measure the full information of a quantum state, process or measurement. Our result can be used for validating Majorana fermion operations~\cite{Irfan2020} and reconstructing processes in fermionic quantum computation systems.

\begin{figure}[tbp]
\centering
\includegraphics[width=1\linewidth]{\figpath 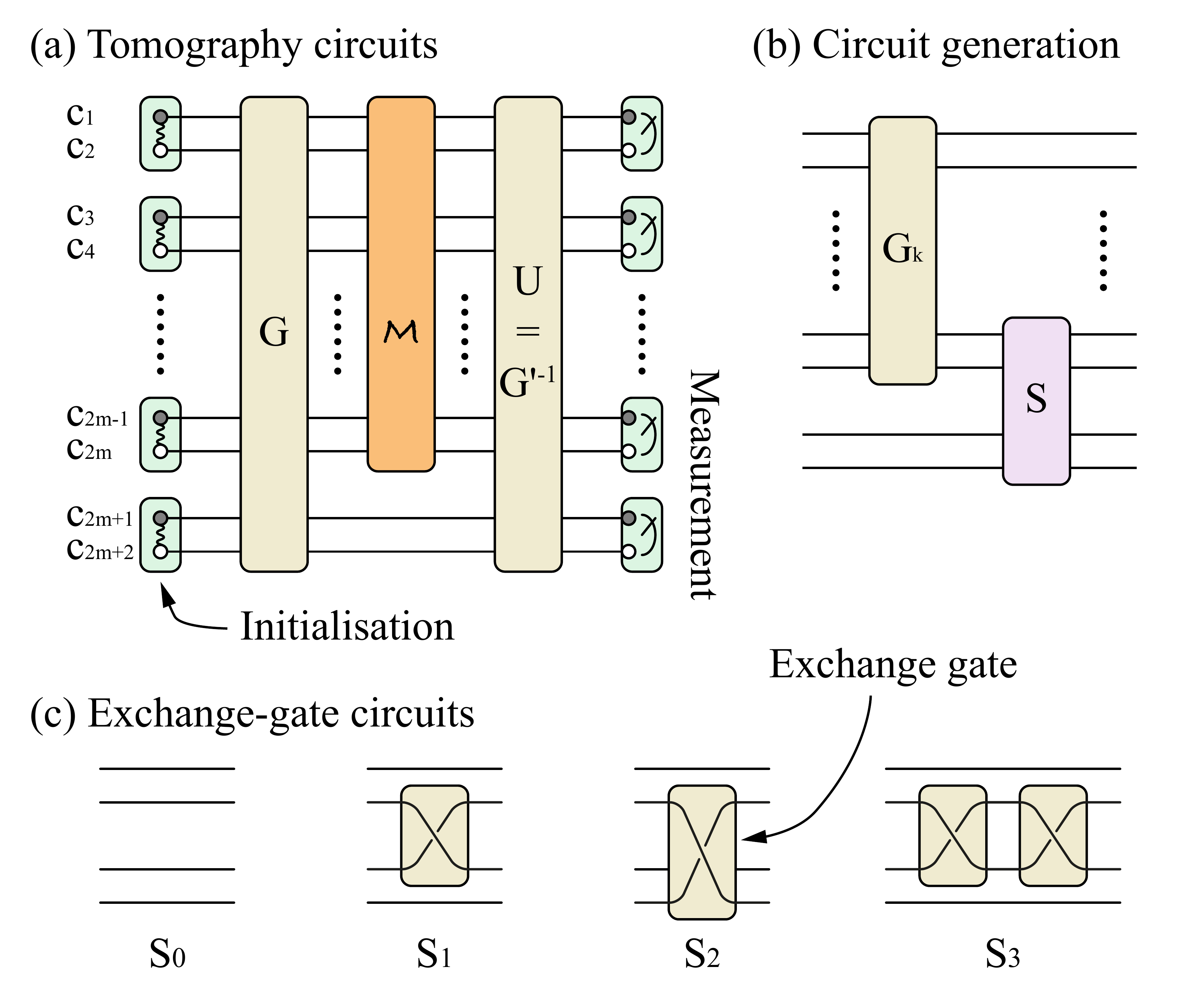}
\caption{
The circuits for the quantum process tomography of Majorana fermion modes. (a) To measure the process $\mathcal{M}$ on $2m$ Majorana fermion modes $c_1,c_2,\ldots,c_{2m}$, we need a pair of ancillary modes $c_{2m+1}$ and $c_{2m+2}$. Majorana fermions are initialised pairwisely in the eigenstate of $ic_{2i-1}c_{2i}$ with the eigenvalue $+1$, where $i = 1,2,\ldots,m+1$. Gates $G,G'\in {\bf G}_{m+1}$ that act on the $2m+2$ modes are generated according to (b) and (c). After the gate $G$, the process $\mathcal{M}$ is implemented on the first $2m$ modes. Following the process $\mathcal{M}$, the gate $U = G^{\prime -1}$ is performed. Then, the final state is measured pairwisely, i.e.~the eigenvalue of each $ic_{2i-1}c_{2i}$ is the measurement outcome.
(b) The circuit for generating ${\bf G}_{m+1}$, see Sec.~\ref{sec:SPprotocol}. The inverse of the circuit generates ${\bf U}_{m+1}$, see Sec.~\ref{sec:Mprotocol}. $G_k$ is a $2k$-mode gate. To generate a $(2k+2)$-mode gate, the $S$ gate is performed on modes $c_{2k-1}$, $c_{2k}$, $c_{2k+1}$ and $c_{2k+2}$. To generate a gate $G$ for the state preparation, we take $S = S_0,S_1,S_2,S_3$; To generate a gate $G'$ for the measurement, we take $S = S_0,S_1,S_2$.
(c)  Circuits built on exchanges gates. Here, $S_0 = \openone$, $S_1 = R_{c_{2k},c_{2k+1}}$, $S_2 = R_{c_{2k},c_{2k+2}}$ and $S_3 = R_{c_{2k},c_{2k+1}}^2$.
}
\label{fig:circuits}
\end{figure}

Fermionic systems obey the superselection rule (SR): The coherence between states with different particle-number parities is forbidden. Therefore, the quantum tomography of fermionic systems have additional restrictions compared with general quantum systems \cite{Amosov2017, Zanardi2002, Banuls2007, Bradler2012}. We first present a set of conditions according to SR that valid quantum states, processes and measurements must satisfy. Then, we introduce the Majorana transfer matrix representation by expressing a density matrix as a linear combination of Majorana fermion operators, which is similar to the Pauli transfer matrix representation for qubit systems. Using this representation, quantum tomography protocols, such as the gate set tomography~\cite{Merkel2013, BlumeKohout2013, Stark2014, Greenbaum2015, BlumeKohout2017, Sugiyama2018}, can be applied to fermionic systems. According to SR, the Majorana transfer matrix of a valid quantum process is block diagonal, and two blocks correspond to transitions between even-parity and odd-parity Majorana fermion operators, respectively. Because valid quantum states are combinations of only even-parity operators, we cannot directly measure the odd block in the process tomography. We show that the odd block can be measured by using one ancillary fermion mode (two Majorana fermion modes).

For a quantum computation system based on Majorana fermions, we will has a finite set of fermion operations. We consider an operation set formed by two-mode initialisation, measurement, unitary gates, and four-mode entangling operations, which is universal for the quantum computation using Majorana fermions~\cite{Bravyi2002}. We find that such an operation set is not even sufficient for measuring the even block without using ancillary modes. For implementing the quantum process tomography in a Majorana fermion quantum computer, we propose tomography circuits based on the universal operation set. The protocol uses two ancillary Majorana fermion modes, the two-mode initialisation, measurement, and exchange gate (the braiding operation). The detailed protocol is illustrated in Fig.~\ref{fig:circuits}.

\section{Formalism of fermionic quantum states, processes and measurements}

Fermions are particles that follow the Fermi-Dirac statistics and Pauli exclusion principle. Discrete fermion modes are described by creation and annihilation operators $a_i^\dag,a_i$ of each mode labelled by $i$, which obey the anticommutation relation $\{ a_i,a_j^\dag \} = \delta_{i,j}$. The vacuum state $\ket{{\rm V}}$ is the state that is empty of particles, i.e.~satisfies $a_j \ket{{\rm V}} = 0$ for all annihilation operators.


Fock states $\ket{\bar{n}} = A_{\bar{n}}^\dag \ket{{\rm V}}$ form an orthonormal basis of the Hilbert space. Here,
\begin{equation}
A_{\bar{n}} = a_1^{n_1} a_2^{n_2} \cdots a_m^{n_m},
\end{equation}
$\bar{n} = ( n_{1}, n_{2}, \ldots , n_{m} )$ is a binary vector, $n_i$ is the occupation number of the mode-$i$ and $m$ is the number of modes. The Hilbert space of the $m$ modes is $\mathcal{H}_m = {\rm span}(\{ \ket{\bar{n}} \})$. Using Fock states, we can explicitly express states, processes and measurements of fermion modes. Fermionic systems obey the superselection rule (SR): The coherence between states with different particle-number parities is forbidden. Compared with general quantum systems, SR introduces additional restrictions on states, processes and measurements of fermion modes. The parity of a Fock state is $(-1)^{\abs{\bar{n}}}$, which is the eigenvalue of the operator $C = \prod_{i=1}^m (\openone-2a_i^\dag a_i)$. Here, $\abs{\bar{n}} = \sum_i n_i$ is the Hamming weight of $\bar{n}$, which is the particle number.

\subsection{State}

The density matrix of fermion modes is an operator on $\mathcal{H}_m$, which can be expressed in the form
\begin{equation}
\rho = \sum_{\bar{n},\bar{n}^{\prime}} \rho_{\bar{n},\bar{n}^{\prime}} \ketbra{\bar{n}}{\bar{n}^{\prime}},
\end{equation}
where $\rho_{\bar{n},\bar{n}^{\prime}}$ are elements of the density matrix. The density matrix must be positive semidefinite and normalised, i.e.~$\rho \geq 0$ and $\Tr(\rho) = 1$. According to SR, $[C, \rho] = 0$, i.e.~$\rho_{\bar{n},\bar{n}^{\prime}} = 0$ for all $\bar{n}$ and $\bar{n}^{\prime}$ that $(-1)^{\abs{\bar{n}}} \neq (-1)^{\abs{\bar{n}'}}$~\cite{Cisneros1998, Hegerfeldt1997, Wick1997, Aharonov1967, Wick1970}.

\begin{theorem}
The density matrix of fermion modes is valid according to SR if and only if $[C, \rho] = 0$.
\end{theorem}

Because $C = C^\dag$ and $C^2 = \openone$, the SR condition can be re-expressed as $\rho = \mathcal{P}(\rho)$, where $\mathcal{P}(\bullet) = \frac{1}{2}\left( \bullet + C\bullet C \right)$. We remark that $\mathcal{P}$ is a projection superoperator, i.e.~$\mathcal{P}^2 = \mathcal{P}$. For all positive semidefinite and normalised density matrices $\rho$, $\mathcal{P}(\rho)$ is a valid state.

\subsection{Process}

The physical process of a quantum system is characterised by a trace-preserving completely-positive map~\cite{Choi1975}. A completely-positive map can be expressed as $\mathcal{M}(\rho) = \sum_q F_q \rho F_q^\dag$, where $F_q$ are Kraus operators on $\mathcal{H}_m$~\cite{Nielsen2010}. The map is trace-preserving if and only if $\sum_q F_q^\dag F_q = \openone$. A map is unital if $\mathcal{M}(\openone) = \openone$. We note that not all physical processes can be described by a completely-positive map, specifically when the system and environment have initial correlations~\cite{Jordan2004}.

\begin{lemma}
The output state of a map $\mathcal{M}$ is valid according to SR for all valid input states if and only if $\mathcal{M}\mathcal{P} = \mathcal{P}\mathcal{M}\mathcal{P}$.
\label{lemma1}
\end{lemma}

According to SR, for the physical process of fermion modes, the output state of the map must be valid for all valid input states. Lemma~\ref{lemma1} can be proved by noticing that $\mathcal{M}\mathcal{P}(\rho) = \mathcal{P}\mathcal{M}\mathcal{P}(\rho)$ for all density matrices $\rho$. However, this is not the sufficient condition of a valid map on fermion modes, similar to the difference between positive and completely-positive maps. For a valid fermionic map, if it acts on a subset of fermion modes, the corresponding composite map on all the modes should also obey SR.

For a system with $m+p$ modes, the Hilbert space is $\mathcal{H}_m\otimes\mathcal{H}_p$, and we use A and B to denote two subsystems, respectively. The Fock state of the composed system reads $\ket{\bar{n},\bar{n}'}_{\rm AB} = \ket{\bar{n}}_{\rm A}\otimes \ket{\bar{n}'}_{\rm B} = A_{{\rm A};\bar{n}}^\dag A_{{\rm B};\bar{n}'}^\dag \ket{{\rm V}}_{\rm AB}$. Note the order of two operators. According to this definition of Fock states for the composite system, we always have $a_{{\rm AB};i} = a_{{\rm A};i}\otimes \openone_{\rm B}$ if the mode-$i$ is in the first $m$ modes, where $a_{{\rm AB};i}$ is the operator that acts on $\mathcal{H}_m\otimes\mathcal{H}_p$, and $a_{{\rm A};i}$ is the operator that acts on $\mathcal{H}_m$. For a mode-$i$ in the other $p$ modes, the operator $a_{{\rm AB};i}$ cannot be written in the tensor product form with the identity operator on the subsystem A, to be consistent with the anticommutation relation. Although fermion operators are not local in general, parity operators are local, i.e.~$C_{\rm A}\otimes\openone_{\rm B}$ is the parity operator of the first $m$ modes, $\openone_{\rm A}\otimes C_{\rm B}$ is the parity operator of the other $p$ modes, and $C_{\rm A} \otimes C_{\rm B}$ is the parity operator of all modes.

The map $\mathcal{M}$ denotes a local process on the first $m$ modes. According to the definition of composite-system Fock states, the map that acts on the composite system can be expressed in the tensor product form $\mathcal{M}_{\rm AB} = \mathcal{M}_{\rm A}\otimes \mathcal{I}_{\rm B}$, where $\mathcal{M}_{\rm A}$ is the map on the operator space of $\mathcal{H}_m$, and $\mathcal{I}_{\rm B}$ is the identity map on the operator space of $\mathcal{H}_p$. We define $\mathcal{P}_{\rm AB}(\bullet) = \frac{1}{2}\left( \bullet + C_{\rm A} \otimes C_{\rm B} \bullet C_{\rm A} \otimes C_{\rm B} \right)$.

\begin{theorem}
A map $\mathcal{M}$ on fermion modes is valid according to SR if and only if
\begin{eqnarray}
(\mathcal{M}_{\rm A}\otimes \mathcal{I}_{\rm B}) \mathcal{P}_{\rm AB} = \mathcal{P}_{\rm AB} (\mathcal{M}_{\rm A}\otimes \mathcal{I}_{\rm B}) \mathcal{P}_{\rm AB},
\end{eqnarray}
for all non-negative integers $p$.
\end{theorem}

Later, we will show that a valid map according to SR is always local.

According to Choi's theorem~\cite{Choi1975}, a map is completely positive if and only if the corresponding Choi matrix is positive semidefinite. For a fermion map $\mathcal{M}$, the Choi matrix is
\begin{eqnarray}
{\rm Choi}(\mathcal{M}) = \mathcal{M}_{\rm A}\otimes\mathcal{I}_{\rm B}\left(\ketbra{\Phi}{\Phi}\right),
\label{eq:Choi}
\end{eqnarray}
where $\ket{\Phi} = \sum_{\bar{n}}\ket{\bar{n}}_{\rm A}\otimes\ket{\bar{n}}_{\rm B}$, and each of A and B has $m$ fermion modes. The map is trace-preserving if and only if $\Tr_{\rm A}\left[{\rm Choi}(\mathcal{M})\right] = \openone_{\rm B}$, and the map is unital if and only if $\Tr_{\rm B}\left[{\rm Choi}(\mathcal{M})\right] = \openone_{\rm A}$.

\begin{theorem}
A map $\mathcal{M}$ on fermion modes is valid according to SR if and only if $[C_{\rm A}\otimes C_{\rm B}, {\rm Choi}(\mathcal{M})] = 0$.
\end{theorem}

The state $\ket{\Phi}$ is an eigenstate of the parity operator $C_{\rm A}\otimes C_{\rm B}$ with the eigenvalue $+1$, i.e.~the total particle number is even. Therefore, if $\mathcal{M}$ is valid,
\begin{eqnarray}
{\rm Choi}(\mathcal{M}) &=& (\mathcal{M}_{\rm A}\otimes\mathcal{I}_{\rm B}) \mathcal{P}_{\rm AB} \left(\ketbra{\Phi}{\Phi}\right) \notag \\
&=& \mathcal{P}_{\rm AB} (\mathcal{M}_{\rm A}\otimes\mathcal{I}_{\rm B}) \mathcal{P}_{\rm AB} \left(\ketbra{\Phi}{\Phi}\right).
\label{eq:PABChoi}
\end{eqnarray}
Because ${\rm Choi}(\mathcal{M}) = \mathcal{P}_{\rm AB}\left( {\rm Choi}(\mathcal{M}) \right)$, we have $[C_{\rm A}\otimes C_{\rm B}, {\rm Choi}(\mathcal{M})] = 0$. The {\it only if} part is proved.

Let $\rho$ be a density matrix of $m+p$ modes. Subsystems A, B and C have $m$ modes, and the subsystem D has $p$ modes.
According to the teleportation formalism,
\begin{eqnarray}
\mathcal{P}_{\rm AD}(\rho_{\rm AD}) = \bra{\Phi}_{\rm BC} \left[ \ketbra{\Phi}{\Phi}_{\rm AB}\otimes \mathcal{P}_{\rm CD}(\rho_{\rm CD}) \right] \ket{\Phi}_{\rm BC}.
\end{eqnarray}
Then,
\begin{eqnarray}
&& (\mathcal{M}_{\rm A}\otimes\mathcal{I}_{\rm D})\mathcal{P}_{\rm AD}(\rho_{\rm AD}) \notag \\
&=& \bra{\Phi}_{\rm BC} \left[ {\rm Choi}(\mathcal{M})_{\rm AB}\otimes \mathcal{P}_{\rm CD}(\rho_{\rm CD}) \right] \ket{\Phi}_{\rm BC}.
\label{eq:teleportation}
\end{eqnarray}
Because $\bra{\Phi}_{\rm BC}\bullet \ket{\Phi}_{\rm BC} = \bra{\Phi}_{\rm BC} \mathcal{P}_{\rm BC}(\bullet) \ket{\Phi}_{\rm BC}$ and $(\mathcal{P}_{\rm BC}\otimes\mathcal{I}_{\rm AD})(\mathcal{P}_{\rm AB}\otimes\mathcal{P}_{\rm CD}) = (\mathcal{P}_{\rm AD}\otimes\mathcal{I}_{\rm BC})(\mathcal{P}_{\rm AB}\otimes\mathcal{P}_{\rm CD})$, we have
\begin{eqnarray}
&& (\mathcal{M}_{\rm A}\otimes\mathcal{I}_{\rm D})\mathcal{P}_{\rm AD}(\rho_{\rm AD}) \notag \\
&=& \bra{\Phi}_{\rm BC} \left[ \mathcal{P}_{\rm AB}({\rm Choi}(\mathcal{M})_{\rm AB})\otimes \mathcal{P}_{\rm CD}(\rho_{\rm CD}) \right] \ket{\Phi}_{\rm BC} \notag \\
&=& \mathcal{P}_{\rm AD}(\mathcal{M}_{\rm A}\otimes\mathcal{I}_{\rm D})\mathcal{P}_{\rm AD}(\rho_{\rm AD}).
\end{eqnarray}
Here, we have used Eq.~(\ref{eq:PABChoi}), Eq.~(\ref{eq:teleportation}) and ($\ket{\Phi}_{\rm BC}$ is an eigenstate of $C_{\rm B}\otimes C_{\rm C}$)
\begin{eqnarray}
&& \bra{\Phi}_{\rm BC} \rho_{\rm ABCD} \ket{\Phi}_{\rm BC} \notag \\
&=& \bra{\Phi}_{\rm BC} (\mathcal{P}_{\rm BC}\otimes\mathcal{I}_{\rm AD})(\rho_{\rm ABCD}) \ket{\Phi}_{\rm BC}.
\end{eqnarray}
The {\it if} part is proved.

A valid state of fermion modes is block diagonal according to the parity. For a valid map $\mathcal{M}$, although off-diagonal blocks of the input and output states are zero, the map may have transitions between elements of off-diagonal blocks. An example is the evolution driven by the Hamiltonian $H = \mu a^\dag a$ for only one mode. The corresponding map is $\mathcal{M}(\bullet) = e^{-i\mu a^\dag a t} \bullet e^{i\mu a^\dag a t}$. These off-diagonal-block transitions do not cause any effect on the output state of the mode. However, the effect can be observed in a multi-mode system, e.g.~the evolution from the state $\frac{1}{\sqrt{2}}(\ket{0,1}+\ket{1,0})$ to $\frac{1}{\sqrt{2}}(\ket{0,1}+e^{-i\mu t}\ket{1,0})$ if the map acts on the first mode.

In general, the parity operator $C$ is not a conserved quantity. For example, the one-mode map $\mathcal{M}(\bullet) = a \bullet a^\dag + a^\dag \bullet a$ is valid and flips the parity. The parity is not conserved when the system and environment exchange particles. The parity is a conserved quantity in unitary processes.

\subsection{Measurement}

The measurement on a quantum system is described by a set of completely-positive maps $\{ \mathcal{E}_k \}$. Each map can be expressed as $\mathcal{E}_k(\bullet) = \sum_q F_{k,q}\bullet F_{k,q}^\dag$, $E_k = \sum_q F_{k,q}^\dag F_{k,q}$ are POVM operators, and $\sum_k E_k = \openone$. Given an input state $\rho$, the probability of the measurement outcome $k$ is $\Tr(E_k\rho)$, and the output state is $\mathcal{E}_k(\rho) / \Tr(E_k\rho)$.

For a measurement on fermion modes, maps $\mathcal{E}_k$ must be valid according to SR. Because $F_{k,q;{\rm A}}\otimes \openone_{\rm B} \ket{\Phi} = \openone_{\rm A} \otimes F_{k,q;{\rm B}}^{\rm T} \ket{\Phi}$, where the transpose is in the Fock basis, we have $E_{k;{\rm B}}^* = \Tr_{\rm A} [{\rm Choi}(\mathcal{E}_k)]$. For a valid map, $[C_{\rm A}\otimes C_{\rm B}, {\rm Choi}(\mathcal{E}_k)] = 0$, then $[C_{\rm B}, E_{k;{\rm B}}^*] = 0$. In the Fock basis, $C_{\rm B}$ is real. Therefore, $[C_{\rm B}, E_{k;{\rm B}}] = 0$, i.e.~$[C, E_k] = 0$.

\begin{theorem}
A set of POVM operators $\{ E_k \}$ on fermion modes is valid according to SR if and only if $[C, E_k] = 0$ for all $k$.
\end{theorem}

The {\it only if} part has been proven. Now we prove the {\it if} part. Given a valid POVM operator $E_k$, there exists a completely-positive map $\mathcal{E}_k(\bullet) = \sqrt{E_k} \bullet \sqrt{E_k}$. Because $E_k$ is positive semidefinte and block diagonal, $\sqrt{E_k}$ is also positive semidefinte and block diagonal. We have $[C, \sqrt{E_k}] = 0$. Then, $\ket{\Psi} = \sqrt{E_{k;{\rm A}}}\otimes\openone_{\rm B} \ket{\Phi}$ is an eigenstate of $C_{\rm A}\otimes C_{\rm B}$ with the eigenvalue $+1$. Because ${\rm Choi}(\mathcal{E}_k) = \ketbra{\Psi}{\Psi}$, $\mathcal{E}_k$ is a valid fermion map.

\section{Majorana fermion operators}

Fermion operators can be written in terms of Majorana fermion operators, $a_i = \frac{1}{2}(c_{2i-1}+ic_{2i})$. Here, $c_i$ are Hermitian (and unitary) operators that obey the anticommutation relation $\{ c_i,c_j \} = 2\delta_{i,j}\openone$. Each fermion mode has to two Majorana fermion operators $c_{2i-1} = a_i + a_i^\dag$ and $c_{2i} = -i(a_i - a_i^\dag)$, i.e.~two Majorana fermion modes.

There are $4^m$ Hermitian operators of $m$ fermion modes in the product form
\begin{eqnarray}
C_{\bar{u}} = i^{\lfloor \abs{\bar{u}}/2 \rfloor} c_1^{u_1} c_2^{u_2} \cdots c_{2m}^{u_{2m}},
\end{eqnarray}
where $\bar{u} = ( u_{1}, u_{2}, \ldots , u_{2m} )$ is a binary vector. These operators have properties similar to Pauli operators:
\begin{itemize}
\item[$\bullet$] Only the trace of $C_{\bar{\bf 0}} = \openone$ is non-zero, i.e.~$\Tr(C_{\bar{u}}) = 2^m\delta_{\bar{u},\bar{\bf 0}}$, where $\bar{\bf 0} = (0, 0, \ldots , 0)$;
\item[$\bullet$] They are Hermitian and unitary, i.e.~$C_{\bar{u}}^2 = \openone$;
\item[$\bullet$] They are orthogonal, i.e.~$\Tr(C_{\bar{u}} C_{\bar{u}'}) = 2^m\delta_{\bar{u},\bar{u}'}$;
\item[$\bullet$] They are commutative or anticommutative with each other, i.e.
\begin{eqnarray}
C_{\bar{u}} C_{\bar{u}'} = (-1)^{\abs{\bar{u}}\cdot\abs{\bar{u}'} + \bar{u}\cdot\bar{u}'} C_{\bar{u}'} C_{\bar{u}},
\end{eqnarray}
where $\bar{u}\cdot\bar{u}'$ is the inner product of two binary vectors~\cite{Bravyi2010};
\item[$\bullet$] $\{ \pm C_{\bar{u}}, \pm i C_{\bar{u}} \}$ is a group, i.e.~$C_{\bar{u}} C_{\bar{u}'} = \eta C_{\bar{u}+\bar{u}'}$, where $\eta = \pm i, \pm 1$, and the $+$ operator of two binary vectors denotes the element-wise summation modulo $2$.
\end{itemize}
Because $ic_{2i-1}c_{2i} = 2a_i^\dag a_i - \openone$, the parity operator $C = (-1)^{m} C_{\bar{\bf 1}}$, where $\bar{\bf 1} = (1, 1, \ldots , 1)$. We can find that $[C, C_{\bar{u}}] = 0$ for all even-parity operators $C_{\bar{u}}$ that $\abs{\bar{u}}$ is even, and $\{C, C_{\bar{u}}\} = 0$ for all odd-parity operators $C_{\bar{u}}$ that $\abs{\bar{u}}$ is odd.

\subsection*{Jordan-Wigner transformation}

Using the Jordan-Wigner transformation, we can express fermion operators using Pauli operators~\cite{Ortiz2001}, which can be used to obtain the explicit expressions of fermion operators in the Fock basis. We decompose the Hilbert space of $m$ fermion modes into $m$ subsystems, i.e.~$\mathcal{H}_m = \mathcal{H}_1^{\otimes m}$, and the Hilbert space of each subsystem is two-dimensional. Accordingly, the Fock state $\ket{\bar{n}} = \bigotimes_i \ket{n_i}$.

The four Pauli operators of one subsystem are $\sigma^I = \ketbra{0}{0} + \ketbra{1}{1}$, $\sigma^X = \ketbra{1}{0} + \ketbra{0}{1}$, $\sigma^Y = i\ketbra{1}{0} - i\ketbra{0}{1}$ and $\sigma^Z = \ketbra{0}{0} - \ketbra{1}{1}$. We define the Pauli operator on $m$ subsystems $S_i = \sigma^{I\otimes (i-1)} \otimes \sigma^S \otimes \sigma^{I\otimes (m-i)}$, which is the operator of $\sigma^S = \sigma^X,\sigma^Y,\sigma^Z$ that acts on the $i$-th subsystem.

In the Fock basis, we can express Majorana fermion operators as~\cite{Gluza2018}
\begin{eqnarray}
&& c_{2i-1} = a_i + a_i^\dag = \sum_{\bar{n}} (-1)^{\sum_{j=i+1}^m n_j} X_i\ketbra{\bar{n}}{\bar{n}}, \\
&& c_{2i} = -i(a_i - a_i^\dag) = \sum_{\bar{n}} (-1)^{\sum_{j=i+1}^m n_j} Y_i\ketbra{\bar{n}}{\bar{n}}.
\end{eqnarray}
Accordingly, the Jordan-Wigner transformation reads
\begin{eqnarray}
&& c_{2i-1} = X_i \prod_{j=i+1}^m Z_j, \\
&& c_{2i} = Y_i \prod_{j=i+1}^m Z_j.
\end{eqnarray}

We remark that the expression in the Fock basis and the Jordan-Wigner transformation depend on the definition of the Fock state, i.e.~the order of fermion operators in $A_{\bar{n}}$, which must be consistent. When using the Fock basis and the Jordan-Wigner transformation, we must take into account all the fermion modes, e.g. including both $m+m$ modes when we compute the Choi matrix.

\section{Majorana transfer matrix}

Similar to the Pauli transfer matrix representation, we can express states, processes and measurements in terms of Majorana fermion operators $C_{\bar{u}}$. These $4^m$ operators are Hermitian and complete (i.e~linearly independent). Therefore, we can always express an operator $F$ as a linear combination of Majorana fermion operators, i.e.
\begin{eqnarray}
F = \sum_{\bar{u}} F_{\bar{u}} C_{\bar{u}}/\sqrt{2^m},
\end{eqnarray}
where $F_{\bar{u}} = \Tr(C_{\bar{u}} F)/\sqrt{2^m}$. If $F$ is Hermitian, coefficients $F_{\bar{u}}$ are real; and $F_{\bar{\bf 0}} = \Tr(F)$.

In the Majorana transfer matrix representation, the state is represented by a $4^m$-dimensional column vector $\HSket{\rho}$ with real elements $\HSket{\rho}_{\bar{u}} = \Tr(C_{\bar{u}} \rho)/\sqrt{2^m}$. Because $\rho$ is normalised, $\HSket{\rho}_{\bar{\bf 0}} = 1/\sqrt{2^m}$. Similarly, a measurement operator is represented by a $4^m$-dimensional row vector $\HSbra{E}$ with real elements $\HSbra{E}_{\bar{u}} = \Tr(C_{\bar{u}} E)/\sqrt{2^m}$.

According to SR, states and measurement operators obey $[C,\rho] = [C,E] = 0$. Therefore, $\HSket{\rho}_{\bar{u}} = \HSbra{E}_{\bar{u}} = 0$ for all $\bar{u}$ with odd $\abs{\bar{u}}$. We define the projections onto two $4^m/2$-dimensional subspaces with even and odd $\abs{\bar{u}}$, receptively, and they are
\begin{eqnarray}
P_{\rm even} &=& \sum_{\bar{u} \st \abs{\bar{u}}\in{\rm Even}}\HSketbra{\bar{u}}{\bar{u}}, \\
P_{\rm odd} &=& \sum_{\bar{u} \st \abs{\bar{u}}\in{\rm Odd}}\HSketbra{\bar{u}}{\bar{u}}.
\end{eqnarray}

\begin{lemma}
In the Majorana transfer matrix representation, for valid states $\rho$ and measurement operators $E$ of fermion modes, the corresponding vectors $\HSket{\rho}$ and $\HSbra{E}$ are in the even subspace, i.e.~$P_{\rm odd}\HSket{\rho} = \HSbra{E} P_{\rm odd} = 0$.
\end{lemma}

\subsection*{Matrix representations of a map}

We can express a completely positive map as
\begin{eqnarray}
\mathcal{M}(\bullet) = \sum_{\bar{u},\bar{u}'} \chi_{\bar{u},\bar{u}'} C_{\bar{u}} \bullet C_{\bar{u}'}.
\end{eqnarray}
The corresponding Choi matrix is
\begin{eqnarray}
{\rm Choi}(\mathcal{M}) = \sum_{\bar{u},\bar{u}'} \chi_{\bar{u},\bar{u}'} C_{\bar{u}} \ketbra{\Phi}{\Phi} C_{\bar{u}'}.
\end{eqnarray}
States $C_{\bar{u}}\ket{\Phi}$ are orthogonal, because $\bra{\Phi} C_{\bar{u}} C_{\bar{u}'} \ket{\Phi} = \Tr(C_{\bar{u}} C_{\bar{u}'})$. The particle-number parity of $C_{\bar{u}}\ket{\Phi}$ is the parity of $\abs{\bar{u}}$. Therefore, for a valid map, $\chi_{\bar{u},\bar{u}'} = 0$ for all $\bar{u}$ and $\bar{u}'$ with different parities, i.e. $\chi$ is block diagonal. Similar to the case of Pauli operators, the map is completely positive if and only if $\chi \geq 0$; the map is trace-preserving if and only if $\Tr(\chi) = 1$.

\begin{theorem}
A map on fermion modes is valid according to SR if and only if $P_{\rm odd} \chi P_{\rm even} = P_{\rm even} \chi P_{\rm odd} = 0$.
\end{theorem}

Now, we can prove that valid maps according to SR are local. We consider four fermion operators $C_{\bar{u}_1}$, $C_{\bar{u}_2}$, $C_{\bar{u}_3}$ and $C_{\bar{u}_4}$ that act on a composite system with $m+p$ modes. In the four operators, $C_{\bar{u}_1}$ and $C_{\bar{u}_2}$ act on the first $m$ modes, and $C_{\bar{u}_3}$ and $C_{\bar{u}_4}$ act on the other $p$ modes. A process on the first $m$ modes has terms in the form $C_{\bar{u}_1}\bullet C_{\bar{u}_2}$; and a process on the other $p$ modes has terms in the form $C_{\bar{u}_3}\bullet C_{\bar{u}_4}$. Note that $\bar{u}_1\cdot \bar{u}_3 = \bar{u}_2\cdot \bar{u}_4 = 0$ because these operators act on different modes. Then, the composite process has terms in the form $C_{\bar{u}_3} C_{\bar{u}_1}\bullet C_{\bar{u}_2} C_{\bar{u}_4} = C_{\bar{u}_1} C_{\bar{u}_3} \bullet C_{\bar{u}_4} C_{\bar{u}_2}$. Here, we have used that $C_{\bar{u}_1}$ and $C_{\bar{u}_2}$ have the same parity, and $C_{\bar{u}_3}$ and $C_{\bar{u}_4}$ have the same parity. Therefore, the process on the first $m$ modes and the process on the other $p$ modes are always commutative.

\begin{theorem}
Two valid maps $\mathcal{M}_1$ and $\mathcal{M}_2$ that act on disjoint modes are always commutative, i.e.~$[\mathcal{M}_1,\mathcal{M}_2] = 0$.
\end{theorem}

The Majorana transfer matrix of a map reads
\begin{eqnarray}
\mathcal{M}^{\rm m}_{\bar{u},\bar{u}'} &=& 2^{-m} \Tr\left[ C_{\bar{u}}\mathcal{M}(C_{\bar{u}'}) \right] \notag \\
&=& 2^{-m} \sum_{\bar{v},\bar{v}'} \chi_{\bar{v},\bar{v}'} \Tr\left( C_{\bar{u}} C_{\bar{v}} C_{\bar{u}'} C_{\bar{v}'} \right).
\end{eqnarray}
The trace is non-zero if and only if $C_{\bar{u}} C_{\bar{v}} C_{\bar{u}'} C_{\bar{v}'} = \eta \openone$, where $\eta = \pm 1,\pm i$ is a phase factor. Therefore, for a non-zero term, $\bar{u}+\bar{v}+\bar{u}'+\bar{v}' = \bar{\bf 0}$. Then, we have $\abs{\bar{u} + \bar{u}'} = \abs{\bar{v} + \bar{v}'}$. Because $\abs{\bar{v} + \bar{v}'}$ is even for all non-zero $\chi_{\bar{v},\bar{v}'}$, $\mathcal{M}^{\rm m}_{\bar{u},\bar{u}'}$ is non-zero only for elements that $\abs{\bar{u} + \bar{u}'}$ is even, i.e.~the Majorana transfer matrix $\mathcal{M}^{\rm m}$ is also block diagonal.

\begin{lemma}
Let $\mathcal{M}^{\rm m}$ be the Majorana transfer matrix of a valid fermion map according to SR. Then $P_{\rm odd} \mathcal{M}^{\rm m} P_{\rm even} = P_{\rm even} \mathcal{M}^{\rm m} P_{\rm odd} = 0$.
\end{lemma}

The map is trace-preserving if and only if $\mathcal{M}^{\rm m}_{\bar{\bf 0},\bar{u}'} = \delta_{\bar{\bf 0},\bar{u}'}$. The map is unital if and only if $\mathcal{M}^{\rm m}_{\bar{u},\bar{\bf 0}} = \delta_{\bar{u},\bar{\bf 0}}$.

\section{Quantum tomography of fermion modes}

With the Majorana transfer matrix representation and the conditions according to SR, we can use conventional quantum tomography protocols to implement the tomography on fermion modes. There are two differences. First, the quantum states, processes and measurements reconstructed in the tomography need to satisfy the SR conditions in addition to other physical conditions. Second, because states and measurements are in the subspaces $P_{\rm even}$, only the even block $\mathcal{M}^{\rm even} = P_{\rm even} \mathcal{M}^{\rm m} P_{\rm even}$ can be directly measured.

In this section, we first discuss the tomography protocols for measuring the even block of a map, and then we prove that the odd block $\mathcal{M}^{\rm odd} = P_{\rm odd} \mathcal{M}^{\rm m} P_{\rm odd}$ can be measured by introducing an ancillary mode (two Majorana fermion modes).

\subsection{Quantum tomography of the even block}

The gate set tomography is a self-consistent process tomography protocol~\cite{Merkel2013, BlumeKohout2013, Stark2014, Greenbaum2015, BlumeKohout2017, Sugiyama2018}, which does not require the prior knowledge on the state preparation and measurement. In this section, we take the gate set tomography as an example. Other quantum tomography protocols, e.g. state tomography and measurement tomography, can be applied to fermionic systems in a similar way.

To implement the gate set tomography, we need to prepare a set of linear-independent states $\HSket{\rho_i}$ and have a set of linearly-independent measurement operators $\HSbra{E_k}$, where $i,k = 1,2,\ldots,4^m/2$. We note that $4^m/2$ is the maximum number of linearly-independent vectors, limited by dimension of the even subspace. These vectors form two $4^m/2$-dimensional matrices $M_{\rm in} = [\HSket{\rho_1}~\HSket{\rho_2}~\cdots~\HSket{\rho_{4^m/2}}]$ and $M_{\rm out} = [\HSbra{E_1}^{\rm T}~\HSbra{E_2}^{\rm T}~\cdots~\HSbra{E_{4^m/2}}^{\rm T}]^{\rm T}$.

The Gram matrix is $g = M_{\rm out} M_{\rm in}$, and each element of the Gram matrix $g_{k,i} = \HSbraket{E_k}{\rho_i} = \Tr(E_k\rho_i)$ can be measured in the experiment, by preparing the state $\rho_i$ and measuring the probability of the measurement operator $E_k$. If $M_{\rm out}$ is known, we can obtain the prepared states by computing $M_{\rm in} = M_{\rm out}^{-1} g$. The state tomography can be implemented in this way, where $M_{\rm in}$ does not need to be a square matrix and can have any number of columns. Similarly, if $M_{\rm in}$ is known, we can obtain measurement operators by computing $M_{\rm out} = g M_{\rm in}^{-1}$. The measurement tomography can be implemented in this way, where $M_{\rm out}$ may not be a square matrix and can have any number of rows.

We consider a set of maps $\{\mathcal{M}_j\}$. The matrix of a map $\mathcal{M}_j$ that can be directly measured in the experiment is $\widetilde{M}_j = M_{\rm out} \mathcal{M}^{\rm even}_j M_{\rm in}$. The element of the matrix is $\widetilde{M}_{j;k,i} = \HSbra{E_k} \mathcal{M}^{\rm m}_j \HSket{\rho_i} = \Tr[E_k \mathcal{M}_j (\rho_i)]$, which can be measured by preparing the state $\rho_i$, implementing the map $\mathcal{M}_j$ and measuring the probability of the measurement operator $E_k$. If $M_{\rm in}$ and $M_{\rm out}$ are know, we can obtain the even block by computing $\mathcal{M}^{\rm even}_j = M_{\rm out} \widetilde{M}_j M_{\rm out}$, which is the process tomography.

In the gate set tomography, if both $M_{\rm in}$ and $M_{\rm out}$ are unknown, we can guess an estimate of $M_{\rm out}$, which is $\widehat{M}_{\rm out}$, and then compute an estimate of $M_{\rm in}$, which is $\widehat{M}_{\rm in} = \widehat{M}_{\rm out}^{-1} g$. The estimate of the even block can be obtained accordingly, which is $\widehat{M}_j = \widehat{M}_{\rm out}^{-1} \widetilde{M}_j \widehat{M}_{\rm in}^{-1}$. These estimates are different from true matrices up to a transformation $T = M_{\rm out}^{-1} \widehat{M}_{\rm out}$: $\widehat{M}_{\rm in} = T^{-1} M_{\rm in}$, $\widehat{M}_j = T^{-1} \mathcal{M}^{\rm even}_j T$ and $\widehat{M}_{\rm out} = M_{\rm out} T$. The gate set tomography is self-consistent in the sense that $\widehat{M}_{\rm out} \widehat{M}_{j_N} \cdots \widehat{M}_{j_2} \widehat{M}_{j_1} \widehat{M}_{\rm in} = M_{\rm out} \mathcal{M}^{\rm even}_{j_N} \cdots \mathcal{M}^{\rm even}_{j_2} \mathcal{M}^{\rm even}_{j_1} M_{\rm in}$ for any sequence of maps.

Such a method for implementing the tomography is called the linear inversion method. An alternative way is based on the maximum likelihood estimation~\cite{Merkel2013, BlumeKohout2013, Stark2014, Greenbaum2015, BlumeKohout2017, Sugiyama2018}. Next, we will discuss how to measure the odd block. For simplicity, we assume that $M_{\rm in}$ and $M_{\rm out}$ are known in the following. The gate set tomography can be directly generalised to the case of measuring the odd block.

\subsection{The full tomography of a fermion process}

To measure the odd block of a map $\mathcal{M}$ that acts on $m$ modes, we introduce an ancillary mode (two Majorana fermion modes). The composite system has $m+1$ modes, the Hilbert space is $\mathcal{H}_m\otimes\mathcal{H}_1$, and we use A and B to denote two subsystems, respectively. Let $\mathcal{M}^{\rm m}_{\rm A} = \mathcal{M}^{\rm even}\oplus \mathcal{M}^{\rm odd}$ be the Majorana transfer matrix of the map on the operator space of $\mathcal{H}_m$. Then, the Majorana transfer matrix of the map on the composite system reads $\mathcal{M}^{\rm m}_{\rm AB} = \mathcal{M}^{\rm even}_{\rm AB}\oplus \mathcal{M}^{\rm odd}_{\rm AB}$, where
\begin{widetext}
\begin{eqnarray}
\mathcal{M}^{\rm even}_{\rm AB} = \left(\begin{array}{cccc}
\mathcal{M}^{\rm even} & 0 & 0 & 0 \\
0 & \mathcal{M}^{\rm odd} & 0 & 0 \\
0 & 0 & \mathcal{M}^{\rm odd} & 0 \\
0 & 0 & 0 & \mathcal{M}^{\rm even}
\end{array}\right)
\begin{array}{c}
C_{\rm even} \\
iC_{\rm odd}c_{2m+1} \\
iC_{\rm odd}c_{2m+2} \\
iC_{\rm even}c_{2m+1}c_{2m+2}
\end{array}
\label{eq:Meven}
\end{eqnarray}
is the even block of the map on the composite system, and
\begin{eqnarray}
\mathcal{M}^{\rm odd}_{\rm AB} = \left(\begin{array}{cccc}
\mathcal{M}^{\rm odd} & 0 & 0 & 0 \\
0 & \mathcal{M}^{\rm even} & 0 & 0 \\
0 & 0 & \mathcal{M}^{\rm even} & 0 \\
0 & 0 & 0 & \mathcal{M}^{\rm odd}
\end{array}\right)
\begin{array}{c}
C_{\rm odd} \\
C_{\rm even}c_{2m+1} \\
C_{\rm even}c_{2m+2} \\
iC_{\rm odd}c_{2m+1}c_{2m+2}
\end{array}
\end{eqnarray}
\end{widetext}
is the odd block of the map on the composite system. Here, the Majorana fermion operators denote the basis the matrix, where $C_{\rm even} = \{ C_{\bar{u}} \st \abs{\bar{u}}\in{\rm Even} \}$ and $C_{\rm odd} = \{ C_{\bar{u}} \st \abs{\bar{u}}\in{\rm Odd} \}$ are $C_{\bar{u}}$ operators on the first $m$ modes with the even and odd parities, respectively.

The even block of the composite system $\mathcal{M}^{\rm even}_{\rm AB}$ can be measured using the quantum tomography, with which we can obtain both $\mathcal{M}^{\rm even}$ and $\mathcal{M}^{\rm odd}$, i.e.~the whole Majorana transfer matrix of the map. If we measure the entire even block of the composite system, we will have two copies of each $\mathcal{M}^{\rm even}$ and $\mathcal{M}^{\rm odd}$. Later, we will show that one can only measure one copy of each block without using more ancillary modes, because of the limited operation set in a Majorana fermion quantum computer.

\section{Quantum tomography in Majorana fermion quantum computers}

The universal quantum computation can be implemented on Majorana fermions with the operation set: i) The preparation of a pair of Majorana fermion modes in the eigenstate of $ic_ic_j$ with the eigenvalue $+1$; ii) The universal gate set including the exchange gate (braiding) $R_{i,j} = \frac{1}{\sqrt{2}}(\openone + c_ic_j)$, the gate $T_{i,j} = e^{\frac{\pi}{8}c_ic_j}$ that enables non-Clifford qubit gates and the entangling gate $\Lambda_{i,j,k,q} = e^{i\frac{\pi}{4}c_ic_jc_kc_q}$; iii) The measurement on a pair of Majorana fermion modes to read out the eigenvalue of $ic_ic_j$~\cite{Bravyi2002}. If the entangling gate is replaced by the four-mode parity projection, the operation set is still universal. The parity projection is a nondestructive measurement of the eigenvalue $c_ic_jc_kc_q$ on four Majorana fermion modes, which is described by two maps $\{ \mathcal{M}_\eta(\bullet) = \frac{\openone +\eta c_ic_jc_kc_q}{2} \bullet \frac{\openone +\eta c_ic_jc_kc_q}{2} \st \eta = \pm 1 \}$; Given the input state $\rho$, when the measurement outcome is the eigenvalue $\eta$, the output state is $\mathcal{M}_\eta(\rho)$ up to a normalisation factor. With the universal operation set, we can implement fault-tolerant universal qubit and fermionic quantum computations~\cite{Li2016,Li2018}. In this section, we discuss how to implement the process tomography in a Majorana fermion quantum computer using the universal operation set.

Given a finite set of fermion modes, we can only measure the even block of a map that acts on these modes. If the total number of Majorana fermion modes is $2m$, the even block is a $4^m/2$-dimensional square matrix. To measure the even block, we need to prepare $4^m/2$ linearly-independent states and have $4^m/2$ linearly-independent measurement operators. However, we are not able to prepare $4^m/2$ linearly-independent states using the limited set of operations in a quantum computer.

First, we consider two Majorana fermion modes, $c_1$ and $c_2$. The two eigenstates of $ic_1c_2$ are $\rho_1 = \frac{1}{2}(\openone + ic_1c_2)$ and $\rho_1' = \frac{1}{2}(\openone - ic_1c_2)$, corresponding to eigenvalues $+1$ and $-1$, respectively. These two states are linearly independent and sufficient for measuring the even block of two Majorana fermion modes. However, usually we only have one initial state in a quantum computer, and let's assume it is $\rho_1$. We can find that the state $\rho_1$ is invariant under gates $S$ and $T$. Therefore, given the initial state $\rho_1$ and the limited set of operations, we cannot measure the entire even block of two Majorana fermion modes.

Now, we consider the general case. For $2m$ Majorana fermion modes, we assume that the initial state of the quantum computer is $\rho_m = \prod_{i=1}^m \frac{1}{2}(\openone + ic_{2i-1}c_{2i})$, which is the eigenstate of all $ic_{2i-1}c_{2i}$ operators with the same eigenvalue $+1$. The state is also an eigenstate of the parity operator with the eigenvalue $(-1)^m$, i.e.~$C \rho_m = \rho_m C = (-1)^m \rho_m$. We can find that the parity operator is a conserved quantity under the four gates $R$, $T$, $\Lambda$ and the parity projection. Therefore, we can only prepare states in a subspace of the Hilbert space, i.e.~states with the parity $(-1)^m$. The dimension of the subspace is $2^{m-1}$, so we can only prepare at most $4^{m-1}$ linearly-independent density matrices using the limited set of operations.

Because of the limited operation set in a Majorana fermion quantum computer, one can only access a subspace of the Hilbert space. To prepare the total $4^m/2$ linearly-independent states, we either need an additional initial state with a different parity or an additional pair of Majorana fermion modes. With the additional modes, we can exchange the particle between two subsystems, which can flip parities of both subsystems. In this way, we can effectively prepare an initial state with the different parity.

In summary, given the limited operation set in a quantum computer, we are not able to measure the entire even block of a map, although it is allowed according to SR. To measure the entire even block, we need a pair of ancillary Majorana fermion modes. Note that we also need the ancillary pair for measuring the odd block. We are about to show that with one ancillary pair, we can measure both even and odd blocks.

\subsection{State preparation protocol}
\label{sec:SPprotocol}


First, we identify the operator space which is the span of states that can be prepared using the limited operation set. We consider $2m$ Majorana fermion modes. The basis of the even-parity operator space is $C_{\rm even}$. All valid states according to SR are in this space. Now, we consider another basis of the space, which is $C_+ \cup C_-$, where $C_\pm = \{ [\openone \pm (-1)^m C]C_{\bar{u}} \st \abs{\bar{u}}\in{\rm Even},~u_{2m}=0 \}$. We note that $(-1)^m CC_{\bar{u}} = \mu_{\bar{u}} C_{\bar{\bf 1}-\bar{u}}$, where $\mu_{\bar{u}} = (-1)^m C C_{\bar{u}}C_{\bar{\bf 1}-\bar{u}} = C_{\bar{\bf 1}} C_{\bar{u}}C_{\bar{\bf 1}-\bar{u}} = \pm 1$ is a function of $\bar{u}$. Basis operators in $C_+$ and $C_-$ are orthogonal and form two $4^{m-1}$-dimensional subspaces. We have $CB = BC = \pm (-1)^m B$ if $B\in C_\pm$. We focus on the operator subspace ${\rm span}(C_+)$.

There are at most $4^{m-1}$ linearly-independent states that can be prepared using the limited operation set. We suppose these states are $\{\rho_i \st i=1,2,\ldots,4^{m-1}\}$, which obey $C \rho_i = \rho_i C = (-1)^m \rho_i$. Therefore, these states form a basis of the subspace ${\rm span}(C_+)$, i.e.~we can express basis operators in terms of states in the form $C_{\bar{u}} + \mu_{\bar{u}} C_{\bar{\bf 1}-\bar{u}} = \sum_i \alpha_{\bar{u},i} \rho_i$, where $\alpha_{\bar{u},i}$ are real coefficients.

Second, we show that $2m+2$ Majorana fermion modes are sufficient for measuring a map on $2m$ Majorana fermion modes. Let $\overline{C}_+$ be the set of basis operators on $2m+2$ Majorana fermion modes that is accessible for the state preparation. We consider four operators
\begin{eqnarray}
G_{\bar{u}} &=& C_{\bar{u}} + \mu_{\bar{u}}C_{\bar{\bf 1}-\bar{u}}ic_{2m+1}c_{2m+2}, \\
H_{\bar{u}} &=& C_{\bar{\bf 1}-\bar{u}} + \mu_{\bar{u}}C_{\bar{u}}ic_{2m+1}c_{2m+2}, \\
I_{\bar{u}} &=& C_{\bar{u}}ic_{2m}c_{2m+1} + \mu_{\bar{u}}C_{\bar{\bf 1}-\bar{u}}c_{2m}c_{2m+2}, \\
J_{\bar{u}} &=& C_{\bar{\bf 1}-\bar{u}}c_{2m}c_{2m+1} - \mu_{\bar{u}}C_{\bar{u}}ic_{2m}c_{2m+2},
\end{eqnarray}
where $\abs{\bar{u}}$ is even and $u_{2m}=0$. We have $\overline{C}_+ = \{ G_{\bar{u}}, H_{\bar{u}}, I_{\bar{u}}, J_{\bar{u}} \}$. Focusing on the first term on RHS of each equation, we can find that $C_{\rm even} = \{ C_{\bar{u}}, C_{\bar{\bf 1}-\bar{u}} \}$ and $C_{\rm odd} = \{ C_{\bar{u}}c_{2m}, -iC_{\bar{\bf 1}-\bar{u}}c_{2m} \}$. In the measurement, if we only measure operators of the first two blocks in Eq.~(\ref{eq:Meven}), i.e.~$C_{\rm even}$ and $iC_{\rm odd}c_{2m+1}$, the second term of each element in $\overline{C}_+$ contributes zero to the measurement result. In this way, we can reconstruct the first two blocks of Eq.~(\ref{eq:Meven}).

Third, we show that operators $G_{\bar{u}}$, $H_{\bar{u}}$, $I_{\bar{u}}$ and $J_{\bar{u}}$ can be constructed using the exchange gate. The exchange gate has the property $R_{i,j}c_iR_{i,j}^\dag = -c_j$ and $R_{i,j}c_jR_{i,j}^\dag = c_i$. We consider the following four states
\begin{eqnarray}
\overline{\rho}_{4i-3} &=& \rho_i\frac{\openone+ic_{2m+1}c_{2m+2}}{2}, \\
\overline{\rho}_{4i-2} &=& R_{2m,2m+1} \overline{\rho}_{4i-3} R_{2m,2m+1}^{-1}, \\
\overline{\rho}_{4i-1} &=& R_{2m,2m+2} \overline{\rho}_{4i-3} R_{2m,2m+2}^{-1}, \\
\overline{\rho}_{4i} &=& R_{2m,2m+1}^2 \overline{\rho}_{4i-3} R_{2m,2m+1}^{-2}.
\end{eqnarray}
Then, we have
\begin{eqnarray}
G_{\bar{u}} &=& \sum_i \alpha_{\bar{u},i} ( \overline{\rho}_{4i-3} + \overline{\rho}_{4i} ), \\
H_{\bar{u}} &=& \mu_{\bar{u}} \sum_i \alpha_{\bar{u},i} ( \overline{\rho}_{4i-3} - \overline{\rho}_{4i} ), \\
I_{\bar{u}} &=& \sum_i \alpha_{\bar{u},i} ( \overline{\rho}_{4i-3} + \overline{\rho}_{4i} - 2\overline{\rho}_{4i-1} ), \\
J_{\bar{u}} &=& \mu_{\bar{u}} \sum_i \alpha_{\bar{u},i} ( \overline{\rho}_{4i-3} + \overline{\rho}_{4i} - 2\overline{\rho}_{4i-2} ).
\end{eqnarray}

Now, we present the state preparation protocol in the inductive way:
\begin{itemize}
\item[1.] Let ${\bf G}_1$ be the set of two-mode gates, and ${\bf G}_1 = \{ \openone \}$ has only one identity gate;
\item[2.] Given the set of $2k$-mode gates ${\bf G}_k$, we generate the set of $(2k+2)$-mode gates [see Fig.~\ref{fig:circuits}(b)~and~(c)], and
\begin{eqnarray}
{\bf G}_{k+1} &=& \{ G, R_{2k,2k+1} G, R_{2k,2k+2} G, \notag \\
&& R_{2k,2k+1}^2 G \st G\in {\bf G}_k \},
\end{eqnarray}
which has $4^{k}$ gates;
\item[3.] Repeat step-2 until ${\bf G}_{m+1}$ is generated;
\item[4.] Prepare the $2m+2$ Majorana fermion modes in the initial state
\begin{eqnarray}
\rho = \prod_{i=1}^{m+1} \frac{\openone + ic_{2i-1}c_{2i}}{2}.
\end{eqnarray}
\item[5.] Generate states $\rho_G = G\rho G^\dag$, where $G\in {\bf G}_{m+1}$.
\end{itemize}
In this way, we can prepare $4^m$ linearly independent states of $2m+2$ Majorana fermion modes, which are sufficient for implementing the process tomography to measure a map on $2m$ Majorana fermion modes.

\subsection{Measurement protocol}
\label{sec:Mprotocol}

We need $4^m$ linearly-independent measurement operators, which are in the form $C_{\rm even}$ and $iC_{\rm odd}c_{2m+1}$. These operators have the even parity and can be written as a product of two-mode operators $ic_ic_j$. Therefore, we can realise the measurement straight-forwardly using two-mode measurements. If we only have two-mode measurements on specific pairs of modes, we can use the exchange gate to realise the two-mode measurement on any pair of modes.

Now, we show that, using pairwise measurements $ic_{2i-1}c_{2i}$ and a subset of inverse gates of $G\in {\bf G}_{m+1}$, we can generate all $4^{m+1}/2$ even-parity operators of the $2m+2$ Majorana fermion modes. For only two Majorana fermion modes, using the measurement of $ic_1c_2$, we can measure two even-parity operators $\{ \openone, ic_1c_2 \}$, where the measurement of the identity operator is trivial. For $2m$ Majorana fermion modes, the set of even-parity operators is $C_{\rm even}$. These even-parity operators can always be measured using pairwise measurements and exchange gates.

For $2m+2$ Majorana fermion modes, even-parity operators can be divided into four subsets [see Eq.~(\ref{eq:Meven})]: $C_{\bar{u}}\in C_{\rm even}$, which are operators without $c_{2m+1}$ and $c_{2m+2}$ in the product; $C_{\bar{u}}i^{1-u_{2m}}c_{2m}c_{2m+1}$, which are operators with $c_{2m+1}$ in the product; $C_{\bar{u}}i^{1-u_{2m}}c_{2m}c_{2m+2}$, which are operators with $c_{2m+2}$ in the product; and $C_{\bar{u}}ic_{2m+1}c_{2m+2}$, which are operators with both $c_{2m+1}$ and $c_{2m+2}$ in the product. We note that all operators in $C_{\rm odd}$ can be expressed as $C_{\bar{u}}i^{-u_{2m}}c_{2m}$, where $C_{\bar{u}}\in C_{\rm even}$.

We show that these four subsets can be measured by applying exchange gates. Given the measurement of $C_{\bar{u}}\in C_{\rm even}$ and the measurement of $ic_{2m+1}c_{2m+2}$, we can measure operators $C_{\bar{u}}$ and $C_{\bar{u}}ic_{2m+1}c_{2m+2}$. If we apply the gate $U^{-1}$ before measurements, we can effectively measure operators $U C_{\bar{u}} U^{-1}$ and $U C_{\bar{u}}ic_{2m+1}c_{2m+2} U^{-1}$. When $u_{2m} = 1$, we have
\begin{eqnarray}
R_{2m,2m+1} C_{\bar{u}} R_{2m,2m+1}^{-1} &=& - C_{\bar{u}}c_{2m}c_{2m+1}, \\
R_{2m,2m+2} C_{\bar{u}} R_{2m,2m+2}^{-1} &=& - C_{\bar{u}}c_{2m}c_{2m+2}.
\end{eqnarray}
When $u_{2m} = 0$, we have
\begin{eqnarray}
&& R_{2m,2m+1} C_{\bar{u}}ic_{2m+1}c_{2m+2} R_{2m,2m+1}^{-1} \notag \\
&=& C_{\bar{u}}ic_{2m}c_{2m+2}, \\
&& R_{2m,2m+2} C_{\bar{u}}ic_{2m+1}c_{2m+2} R_{2m,2m+2}^{-1} \notag \\
&=& - C_{\bar{u}}ic_{2m}c_{2m+1}.
\end{eqnarray}
Therefore, all operators in the four subsets can be measured.

Similar to the state preparation, we present the measurement protocol in the inductive way:
\begin{itemize}
\item[1.] Let $U_1$ be the set of two-mode gates, and ${\bf U}_1 = \{ \openone \}$ has only one identity gate;
\item[2.] Given the set of $2k$-mode gates ${\bf U}_k$, we generate the set of $(2k+2)$-mode gates, and
\begin{eqnarray}
{\bf U}_{k+1} &=& \{ U, U R_{2k,2k+1}^{-1}, G R_{2k,2k+2}^{-1} \st U\in {\bf U}_k \};
\end{eqnarray}
\item[3.] Repeat step-2 until ${\bf U}_{m+1}$ is generated;
\item[4.] At the end of the circuit, implement the pairwise measurements of $\{ ic_{2i-1}c_{2i} \st i=1,2,\ldots,m+1 \}$. With these pairwise measurements, we can effectively measure $2^{m+1}$ operators in the form
\begin{eqnarray}
Q_{\bar{n}} = \prod_{i=1}^{m+1} (ic_{2i-1}c_{2i})^{n_i}.
\end{eqnarray}
\item[5.] Apply the gate $U\in {\bf U}_{m+1}$ before the measurement to effectively measure the operator $U^\dag Q_{\bar{n}}U$. The set $\{ U^\dag Q_{\bar{n}}U \}$ includes all even-parity operators of the $2m+2$ Majorana fermion modes.
\end{itemize}
We can find that $U^{-1} \in {\bf G}_k$ for all $U \in {\bf U}_k$.

\subsection{The protocol of tomography}

The process tomography of the map $\mathcal{M}$ on $2m$ Majorana fermion modes can be measured as shown in Fig.~\ref{fig:circuits}:
\begin{itemize}
\item[1.] To measure the map on $2m$ modes, we need two ancillary Majorana fermion modes;
\item[2.] Generate the gate set ${\bf G}_{m+1}$ according to the circuit in Fig.~\ref{fig:circuits}(b)~and~(c), and circuits are generated using four gates $S = S_0,S_1,S_2,S_3$;
\item[3.] Generate the gate set ${\bf U}_{m+1}$ formed by inverse gates $U = G^{-1}$, where $G\in {\bf G}_{m+1}$ are generated by three gates $S = S_0,S_1,S_2$;
\item[4.] Prepare each pair of Majorana fermion modes in the eigenstate of $ic_{2i-1}c_{2i}$ with the eigenvalue $+1$;
\item[5.] Choose a gate $G\in {\bf G}_{m+1}$ and apply the gate on $2m+2$ modes;
\item[6.] Apply the map on the first $2m$ modes;
\item[7.] Choose a gate $U\in {\bf U}_{m+1}$ and apply the gate on $2m+2$ modes;
\item[8.] Measure the operator $ic_{2i-1}c_{2i}$ for each pair of Majorana fermion modes.
\end{itemize}
Data generated using these circuits are informationally complete and sufficient for computing both the even and odd blocks of the map.

The state preparation and measurement circuits are simulated numerically using the code available at \href{https://github.com/Zhanggangtjnu/FMQPT.git}{https://github.com/Zhanggangtjnu/FMQPT.git}.

\section{Conclusions}

In this paper, we present the conditions that valid fermion states, processes and measurements must satisfy according to SR. We introduce the Majorana transfer matrix representation, such that tomography protocols can be applied to fermionic systems. According to SR conditions, we find that the full reconstruction of fermion processes always needs at least one ancillary fermion mode (two Majorana fermion modes). In a Majorana fermion quantum computer, the informationally-complete state preparation and measurement can be realised using two-mode initialisation, measurement and braiding operations, and the protocol is explicitly presented. Our results can be used for the validation of quantum operations and reconstruction of processes in Majorana fermion systems.

\acknowledgments
We thank Wenyu Wu for discussions on the case with particle number conservation.
This work is supported by the National Natural Science Foundation of China (Grant No. 11705127, 11574028, 11874083 and 11875050).
GZ acknowledges the support of Program for Innovative Research in University of Tianjin (Grant No. TD13-5077).
YL acknowledges the support of NSAF (Grant No. U1930403).

\end{document}